\documentclass[5p,times,numbers,square]{elsarticle}
\usepackage{amsmath,amssymb,bm}
\usepackage{graphicx}	            	
\usepackage[hidelinks]{hyperref}    	

\begin{document}
\title{Geodesic Noise and Gravitational Wave Observations by Pulsar Timing Arrays}
\author[1]{Sebastian Golat}
\ead{s.golat19@imperial.ac.uk}
\author[1]{Carlo R. Contaldi}
\address[1]{Blackett Laboratory, Imperial College London, SW7 2AZ, UK}

\date{\today}
\begin{abstract}
Signals from millisecond pulsars travel to us on geodesics along the line-of-sight that are affected by the space--time metric . The exact path-geometry and redshifting along the geodesics determine the observed Time-of-Arrival (ToA) of the pulses. The metric is determined by the distribution of dark matter, gas, and stars in the galaxy and, in the final stages of travel, by the distribution of solar system bodies. The inhomogeneous distribution of stellar masses can have a small but significant statistical effect on the ToAs through the perturbation of geodesics. This will result in additional noise in ToA observations that may affect Pulsar Timing Array (PTA) constraints on gravitational waves at very low frequencies. We employ a simple model for the stellar distribution in our galaxy to estimate the scale of both static and dynamic sources of what we term generically ``geodesic noise''. We find that geodesic noise has a standard deviation of ${\cal O}(10)$~ns for typical lines-of-sight. This suggests geodesic noise is relevant for estimates of PTA sensitivity and may limit future efforts for detection of gravitational waves by PTAs.
\end{abstract}

\maketitle
\section{Introduction}%
Many pulsars are well known to be a very precise celestial clocks due to their rotational stability. This makes them an excellent tool for examining phenomena that can affect the pulse's Time-of-Arrival (ToA) (see for example~\cite{Lorimer2012}). Gravitational waves (GWs) will affect the ToA by perturbing the length of geodesics along which pulsar signals travel, inducing time delays. This effect is potentially observable \cite{Hulse1975}. In particular, a GW crossing through multiple pulsar lines-of-sight will induce a correlated delay in the ToAs of each pulsar \cite{Hellings1983}. The effect is relatively weak and other effects will affect the ToA \cite{Lorimer2012}. Therefore, one needs high precision timing. This was made possible thanks to the discovery of millisecond pulsars in 1982 by \citet{Backer1982}. The suggestion to correlate timing data from a network of millisecond pulsars emerged in the 1990s \cite{Foster1990}. These ideas eventually evolved into dedicated programmes known as Pulsar Timing Arrays (PTAs). During the last decade, there has been a global effort to use these to detect GWs. Some of the current PTAs are North American Nanohertz Observatory for Gravitational Waves (NANOGrav) \cite{Jenet2009}, European Pulsar Timing Array (EPTA) \cite{Kramer2013}, The Parkes Pulsar Timing Array (PPTA) \cite{Hobbs2013}, and the combined International Pulsar Timing Array (IPTA) \cite{Hobbs2009}. It is a common belief that within a decade, PTAs should be able to detect GWs with a period of the order of years or frequencies ${\cal O}(10^{-8})$~Hz \cite{Lommen2015}.

The analysis of ToA data from pulsars includes various deterministic corrections to the raw data to account for a number of effects. One of these is the Shapiro delay experienced by the signal as it travels through the solar system. The delay is due to the gravitational potential of bodies in our solar system which distorts geodesics and induces an increase in the path length travelled by the signal. The Shapiro delay was first considered as a test of the general theory of relativity by~\citet{Shapiro1964}. A more generalised treatment of this geodesic delay in the weak gravitational field of arbitrary-moving bodies was presented in \citet{Kopeikin1999}. While the delay due to the gravitational potential of solar system bodies is accounted for \cite{Edwards2006} there has been no discussion of the {\sl statistical} effect due to the distribution of massive bodies in our galaxy. We may naively expect this signal to be too small to be relevant. However, the amplitude of the GW signal being searched for is also very small, on the order of nanoseconds \cite{Hobbs2006}. The overall delay due to the gravitational potential well of our galaxy is of the order of days for typical pulsar--Earth lines-of-sight \cite{Desai2015}. Even small perturbations to this overall effect may be resolvable. A small, but significant, fraction of the total mass of our galaxy is made up of compact sources---effectively point sources compared to the typical pulsar--Earth distances. We should not expect delays due to individual stars to be relevant, the chances of a pulsar line-of-sight passing sufficiently close to an individual star are negligible. However, one may worry that a distribution of sources may lead to mean field effects that counter the inverse distance scaling of individual potentials. This statistical contribution to ToA delays would be small but perhaps non-negligible for individual lines-of-sight. Furthermore, this effect, in general, is {\sl dynamical} since each pulsar has a peculiar velocity with respect to Earth, as do all the stars contributing to the potential. Consequently, the line-of-sight geodesic deformation will change over time.

In this {\sl letter} we make a first attempt at estimating the size of this additional contribution to ToA uncertainty which we term ``geodesic noise''. We briefly review how timing residuals are fitted in pulsar observations and then consider the generalised time delay induced by a distribution of point masses---this is analogous to the generalised Coulomb problem in electromagnetism \cite{Jackson1998} with a particular analogy to how charges move through a conductor in the presence of random Coulomb sources. We then consider the application to a very simplified model of point-source distributions in our galaxy from which we sample time delays over a distribution of lines-of-sight. Our results show that the effect is non-negligible and gives a noise contribution at a level that is comparable to the timing accuracy of next-generation PTA efforts such as the Square Kilometre Array (SKA) \cite{Moore2014}.

\section{Pulsar timing residuals}%
The standard procedure in pulsar timing is to fit the ToA of each pulsar with a quadratic timing model \cite{Moore2015}
\begin{equation}\label{eq:model}
    m^I(t_k)=\alpha^I+\beta^I t^{\,}_k +\gamma^I t_k^2\,,
\end{equation}
where capital letter superscripts $I$, $J$ label the individual pulsars and $t_k$ are times of observations at different epochs. The cadence of observation epochs is typically fortnightly (see e.g. \cite{Moore2015}) and the ToA at each epoch $t_k$ is itself obtained by a fit of template or model profiles to an average pulse profile that is obtained by averaging over many pulses. The fit estimates the pulsar's phase evolution; $\alpha^I$ represents a constant phase offset, $\beta^I$ is related to the pulsar's rotational frequency and peculiar velocity, and $\gamma^I$ is related to the spin-down rate and peculiar acceleration \cite{Moore2015}. The difference between the ToA and timing model of each pulsar is the timing residual. The cross-correlation between pulsar residuals $\langle R^I_k R^J_k\rangle$ contains a particular dependence on the cosine of the angle between the lines-of-sight $\bm{\hat n}^I\cdot\bm{\hat n}^J$ if a GW has passed through during the observation period.

Noise in the timing of the millisecond pulse profiles and errors in the solar system Shapiro delay modelling induce an effective error in the residual on top of any signal $S_I$
\begin{equation}
    R^I_k = S^I_k + \epsilon^I_k\,,
\end{equation}
where $\epsilon_k$ can be modelled as Gaussian, mean free, random variate with variance\footnote{although the $\delta^{IJ}$ is an assumption that may break down due to correlated errors in the solar system modelling} $\langle \epsilon^I_k \epsilon^J_{k'}\rangle = \delta^{IJ}\delta_{kk'}\sigma^I_k$. Since the error, to a first approximation, is uncorrelated between pulsars it will not bias the ensemble mean of cross-correlations in the residuals but does contribute to the variance about their means. It is, therefore, a source of variance in searches for GWs using PTA observations. Typical values for $\sigma_k$ for current generation of PTAs is $\sigma_k\sim{\cal O}(100)$~ns \cite{Moore2015}.

If any delay arising from outside the solar system was constant, or if any dynamical contribution were solely due to the peculiar motion of the pulsar, PTA observations would not be affected since these contributions would be fitted out when applying the timing model (\ref{eq:model}). This assumption would hold if the gravitational potential were solely determined by an isotropic, smooth distribution of matter. This is more or less the case for the dark matter halo that dominates the potential of our galaxy and, to a lesser extent, the diffuse baryonic matter that makes up the next biggest contribution. It is certainly not the case for the contribution from the matter contained in stars, each of which also contributes through a peculiar velocity on top of the coherent, rotational motion of the galaxy. The distortions of geodesics due to the distribution of stars in the galaxy will source another contribution to the residual error $\eta^I_k$ that is also uncorrelated, in time, between pulsar lines-of-sight but may contain an overall dependence on orientation.

\section{Stellar delay modelling}%
The gravitational potential due to stars in the galaxy can be modelled as a sum over individual contributions from point sources
\begin{equation}
    \phi(\bm{r}) = \sum_i \frac{GM_i}{|\bm{r}-\bm{r}_i|}\,,
\end{equation}
where $i$ runs over all individual stars in the galaxy each with mass $M_i$ and located at position $\bm{r}_i$. The Shapiro time delay is given by the integral of the potential along the geodesic
\begin{equation}
    \Delta =\frac{2}{c} \int \phi(\bm{r})\,\mathrm{d}\bm{\ell}\,,
\end{equation}
where $c$ is the speed of light and $\mathrm{d}\bm{\ell}$ is the infinitesimal line element along the geodesic.

The generalisation of the time delay due to a distribution of moving point masses has been studied by \citet{Kopeikin1999} who obtain a covariant expression for the time delay observed at a location $\bm{r_\mathrm{o}}$ at retarded time associated with proper time $t_\mathrm{o}$ for a photon with momentum at past null infinity aligned with unit vector $\bm{\hat k}$ (for more details see \cite{Kopeikin1999})
\begin{equation}\label{eq:timedelay}
    \Delta(\bm{r}_\mathrm{o}, t_\mathrm{o})=-\frac{2 G}{c^{3}} \sum_i M_i\ln \left[\frac{\left(r_i-\bm{\hat k} \cdot \bm{r}_i\right)^{1-\bm{\hat k} \cdot \bm{v}_i/c}}{\left(r_{\mathrm{o}i}-\bm{\hat k} \cdot \bm{r}_{\mathrm{o}i}\right)^{1-\bm{\hat k} \cdot \bm{v}_{\mathrm{o}i}/c}}\right]\,.
\end{equation}
Here $\bm{r}_{\mathrm{o}i}=\bm{r}_\mathrm{o}-\bm{r}_i$, $\bm{v}_{\mathrm{o}i}=\bm{v}_\mathrm{o}-\bm{v}_i$, and $\bm{v}_\mathrm{o}$ and $\bm{v}_i$ are the velocity vectors of the observer and point mass $M_i$ respectively. The expression (\ref{eq:timedelay}) neglects a sub-dominant term due to the accelerations of the masses which involves an integral over retarded times and is therefore difficult to calculate. The positions and velocities in (\ref{eq:timedelay}) are also sampled at retarded times but we neglect this in order to significantly speed up the calculation. The effect of not sampling the distribution using retarded time at every mass location is not thought to be significant since this is a statistical realisation of the stellar distribution. Another approximation is that we take the unit vector $\bm{\hat k}$ to be the direction from the pulsar to the observer in flat space-time.

\begin{table}[t]%
    \centering
    \caption{Thin and thick disk parameters for the Galactic model (\ref{eq:density}). }\label{tab:galaxy}
    \begin{tabular}{c c c r}
    \hline
    Parameter&Thin &  Thick &\\
    \hline
        $z_\text{d}$& $0.3$&$0.9$&$\text{kpc}$\\
        $R_\text{d}$& $2.6$&$2.0$&$\text{kpc}$\\
        $M_\text{d}$& $3.5$& $6.0$&$\times\,10^{10}M_\odot$\\
    \hline
    \end{tabular}
\end{table}%

We apply the time delay to a simple stellar model of the Galaxy. The model consists of masses $M_i$ located at centres of a Cartesian grid. Our typical simulation contains $8$ million cells with some $4.4$ million cells containing nonzero mass. These masses are drawn from a Poisson distribution with mean $\langle M_i\rangle=\rho_{\mathrm{gal}}(R, z)\Delta{x}\Delta{y}\Delta{z}$, where $R$, $\theta$, and $z$ are cylindrical coordinates and $\rho_{\mathrm{gal}}(R, z)$ is the mass density distribution for stars in the Galaxy. The density is made up of two distinct contributions; one for a thin disc and another for a thick disc \cite{McMillan2016}, each having the form
\begin{equation}\label{eq:density}
\rho_{\mathrm{d}}(R, z)=\frac{\Sigma_{d}}{2 z_{\mathrm{d}}} \exp \left(-\frac{|z|}{z_{\mathrm{d}}}-\frac{R}{R_{\mathrm{d}}}\right)\, , 
\end{equation} 
where $z_{\mathrm{d}}$ and $R_{\mathrm{d}}$ are scale heights and lengths respectively and $\Sigma_{d}=M_{\mathrm{d}}/(2 \pi R_{\mathrm{d}}^{2})$ is a surface density with $M_{\mathrm{d}}$ the total disc mass. We adopt values for the thin and thick disc parameters from \citet{Binney2008} (see table~\ref{tab:galaxy}). We also assign peculiar velocity components for each mass cell by drawing from a normal distribution with zero mean and standard deviation $20\,\text{kms}^{-1}/\sqrt{3}$ such that the root mean square speed is the deviation of circular speed measured in the neighbourhood of our galaxy \cite{Binney2008}.

\begin{figure}[t]
    \centering
    \includegraphics{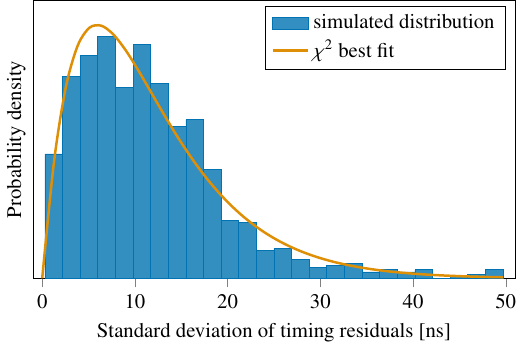}
    \caption{Histogram of geodesic noise standard deviation $\eta$ obtained using a stellar model grid with cell size $0.2\times0.2\times0.1\,\text{kpc}^3$. The histogram is normalised as a probability density. The number of lines-of-sight is $N_\text{p}={1000}$. The solid curve shows a $\chi^2$ fit to the distribution with median $\eta = 9.8$~ns and $95\%$ upper limit from the continuous distribution is given by percentile is $\eta < 27.5$~ns. The distribution of ToA delays is significantly non-Gaussian.}
    \label{fig:density}
\end{figure}

{\em Geodesic noise.---}%
The stellar delay model discussed above can be used to calculate the contribution from geodesic noise for individual pulsars ToAs. To do this we draw random locations for $N_\text{p}$ pulsars located uniformly at radial distance $R\in\left[0.15,10\right]$~{kpc}, at azimuth $\theta\in\left[0,2\pi\right]$ and axial coordinate $z\in\left[-300,300\right]$~pc. We integrate the delay (\ref{eq:timedelay}) along each line-of-sight and repeat this for a total time of 15 years at fortnightly intervals to model the typical cadence of PTA observations. We then fit the timing model (\ref{eq:model}) to these delays in order to calculate the additional residuals that are due to the geodesic noise.

\begin{figure}[t]
    \centering
    \includegraphics{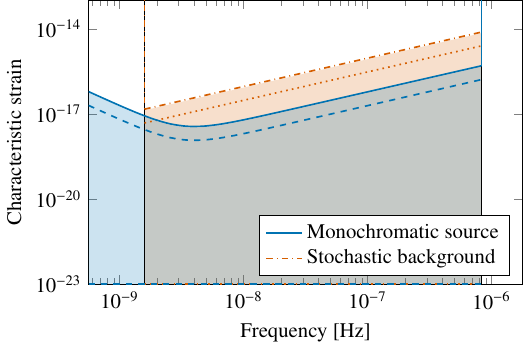}
       \caption{Impact of geodesic noise on SKA GW sensitivity curves. For the purpose of this estimate we have assumed that SKA will monitor 100 pulsars timed with a fortnightly cadence for 20 years with timing standard deviation of $30\,\text{ns}$. The blue (solid) line is the sensitivity to monochromatic sources (\ref{eq:mono}) and red (dash-dotted) line is for a stochastic background (\ref{eq:gwb}). The blue (dashed) and red (dotted) lines show the additional contribution from the median value of $\eta$ for both cases respectively. The corresponding shaded regions show the 95\% upper limit on geodesic noise - these are very close to the raw timing sensitivity curves.}
    \label{fig:sensitivity}
\end{figure}

We calculate the sample standard deviation of these residuals from the sequence of fortnightly delays in order to estimate the stochastic contribution $\eta$ to the noise of individual pulsars. We find that the static contribution of the total delay dominates over the velocity contribution. This is not surprising since the typical peculiar velocities of the stars have $v\ll c$. Figure~\ref{fig:density} shows the histogram for $\eta$ obtained using $N_{\text{p}}=1000$ pulsar lines-of-sight observed for 15 years with a fortnightly cadence. The histogram is significantly non-Gaussian and is well approximated by a $\chi^2$ distribution with median $\eta=9.8$~ns and 95\% upper limit $\eta < 27.5$~ns.

We leave a detailed exploration of the dependence of this distribution on the stellar model parameters and simulation resolution for future work, but we have verified that the result is stable with respect to the grid resolution. The peculiar velocity distribution for stars in the galaxy is not well understood and this is a significant source of uncertainty in our modelling but since the dynamical component to due terms $\bm{v}_i$ in (\ref{eq:timedelay}) is sub-dominant to the static one we do not expect that this will have a strong effect on the results. Our uniform distribution of lines-of-sights may also affect the details of the distribution, it is not clear how observed millisecond pulsars are distributed in the Galactic volume as there are only a small number being used in current PTA networks. Pulsar locations should be drawn from a distribution that is weighted by the stellar density. This would lead to more lines-of-sight through the Galactic disc. Intuitively, we expect this would increase the mean value of $\eta$ but reduce the number of outliers. This would have the effect of making the distribution of $\eta$ more Gaussian.

\section{Effect on GW observations}%
Geodesic noise is an additional contribution in PTA GW searches. We can estimate the impact on GW searches by using our estimate of $\eta$ as the timing noise in sensitivity calculations for the characteristic strain $h_c$ observed by PTA networks as a function of frequency $f$.
The sensitivity curves can be calculated for two distinct cases. The first is for the case of a monochromatic source \cite{Moore2015}
\begin{equation}
h_\mathrm{c}(f)\approx\left(\frac{432}{{N}_{\mathrm{p}}\left({N}_{\mathrm{p}}-1\right)}\right)^{\frac{1}{4}}\eta \sqrt{\frac{\delta{t}}{{T}}}\left({f}+\frac{8}{f^2T^3}\right) ,\label{eq:mono}
\end{equation}
where $N_\text{p}$ is the number of pulsars in the array, $1/\delta{t}$ is the cadence and $T$ is the overall length of observation.
This would be relevant for GWs arising from individual astrophysical systems such as super-massive black hole binaries. The second is the case of a stochastic background \cite{Moore2015}
\begin{equation}
h_\mathrm{c}(f)\approx \frac{700 }{\sqrt{{N}_{\mathrm{p}}\left({N}_{\mathrm{p}}-1\right)}}\eta\sqrt{\frac{\delta{t}}{{T}}}f .\label{eq:gwb}
\end{equation}
 This would arise for the case where the signal is made up of many confused sources or, in principle, a cosmological background generated at high redshift. We use the same normalisation factors as obtained by \citet{Moore2015} in (\ref{eq:mono}) and (\ref{eq:gwb}). In Figure~\ref{fig:sensitivity} we show how the median and 95\% upper limit on $\eta$ translates into characteristic strain sensitivity for both cases. We compare this with the sensitivity curves for SKA assuming $T=20$ years monitoring $N_\mathrm{p}= 100$ pulsars timed fortnightly with an expected timing error standard deviation of $30$~ns \cite{Moore2014}.

\section{Discussion}%
We have examined how the deformation of pulsar signal geodesics by stars in the galaxy may introduce a significant source of additional noise in PTA observations aimed at detecting low-frequency GWs. Our initial, albeit simplistic, modelling of the effect indicates that the ``geodesic noise'' induced is of an order comparable to the raw timing noise limiting the next generation of PTA efforts such as that planned for SKA. It will, therefore, have an impact on the overall sensitivity for future PTA efforts and should be considered in such estimates. We have carried out a simple analysis of the impact of geodesic noise on SKA sensitivity curves. In principle, geodesic noise will itself have non-trivial angular correlations which may help in mitigating its effect, but this would only be possible in the limit of large numbers of observed pulsars. Even surveys such as SKA expect to be able to survey only on the order of a few hundred millisecond pulsars with sufficient timing stability for GW searches. 

The frequency dependence of PTA signals will be useful in assessing whether any observed signal is due to this effect. Geodesic noise will appear as additional Poisson noise in PTA data. This implies a flat spectrum which can be characterised in the data to reject any hypothesis that the signal power may be due to geodesic noise. 

Geodesic noise is distinct from the noise induced by interstellar dispersion variability \cite{Lorimer2012}. This arises due to the presence of ionised gas clouds along the line-of-sight. The dispersion effect is frequency dependent and can be modelled accurately given sufficiently precise data and frequency coverage.

A natural question to ask is whether geodesic noise due to the stellar distribution is already affecting ToAs at the current level of timing resolution. It is well known that millisecond pulsars display intrinsic timing instabilities on long timescales \cite{Verbiest_at_al2015}. This noise is particularly important for low-frequency GW searches and if it is driven by geodesic noise it may be useful to study the effect in more detail. For example, a detailed understanding of the noise spectrum induced by this effect may help to mitigate its impact on PTA analysis.

Another interesting question is whether the contribution from any inhomogeneity in the dark matter distribution can be dismissed as was discussed in \citet{Baghram2011}. If the dark matter distribution has significant sub-structure on galactic scales its contribution to geodesic noise may be significant or dominant. Millisecond pulsar ToA observations may, therefore, lead to significant constraints on such models of dark matter.

\section*{Acknowledgement}
CRC acknowledges support under a UKRI Consolidated Grant ST/T000791/1.  SG is supported by a UKRI Studentship under grant ST/T506151/1.

\hbadness=10000\relax

\end{document}